\documentstyle[pre,preprint,aps]{revtex}
\tightenlines
\begin{document}
\draft
%\baselineskip=0.5\baselineskip

\title{
Dynamical mean-field approximation to \\
coupled active rotator networks
subject to white noises
%\footnote{E-print: cond-mat/0206135}
}
\author{
Hideo Hasegawa
\footnote{E-mail:  hasegawa@u-gakugei.ac.jp}
}
\address{
Department of Physics, Tokyo Gakugei University,
Koganei, Tokyo 184-8501, Japan
}
\date{\today}
\maketitle
\begin{abstract}
A semi-analytical
dynamical mean-field approximation (DMA) has been developed
for large but finite $N$-unit active rotator (AR) networks
subject to individual white noises.
Assuming weak noises and the Gaussian 
distribution of state variables, 
we have derived equations of motions for moments of 
local and global variables up to
the {\it infinite} order.
In DMA, the original $N$-dimensional 
{\it stochastic} differential equations (DEs) 
are replaced by three-dimensional {\it deterministic} DEs
while the conventional moment method yields
$(1/2)N(N+3)$ deterministic DEs for 
moments of local variables. We have discussed
the characters of the stationary state, the time-periodic state
and the random, disordered state, which are realized 
in excitable AR networks depending
on the model parameters. 
It has been demonstrated that 
although fluctuations of global variable
vary as $1/\sqrt{N}$ when $N$ is increased, 
those of local variables remain finite 
even for $N \rightarrow \infty$. 
Results calculated with the use of our DMA
are compared to those obtained by direct simulations
and by the Fokker-Planck equation which is applicable
to the $N=\infty$ AR model. 
The advantage and disadvantage of DMA are also discussed.

\end{abstract}

%\vspace{0.5cm}
%\pacs{PACS No. 05.45.-a 05.45.Xt 05.70.Fh 87.10.+e}

\vspace{1.0cm}

%\noindent

%\vspace{1.0cm}
%\noindent
%Correspondence to the author:  

%Prof. Hideo Hasegawa, 

%Dep. of Physics, Tokyo Gakugei Univ.,
%4-1-1 Nukin-kita machi, Koganei, Tokyo 184-8501

%Tel: 042-329-7482; Fax: 042-329-7491

%E-mail: hasegawa@u-gakugei.ac.jp

\newpage
%\narrowtext
\section{INTRODUCTION}

%\begin{center}
%{\bf I. INTRODUCTION}
%\end{center}
It has been shown that noises play intrigue and essential
roles in non-linear systems. 
Some examples are the noise-induced transition \cite{Horst84}
and the stochastic resonance \cite{Gammai98}.
Effects of noises 
on the dynamics of a variety of real systems such as 
the Josephson junction \cite{Kim93}, chemical reaction \cite{Kuramoto84}, 
charge density of states \cite{Fisher85} 
and neural networks \cite{Sompo91},
have been studied with the use of 
the coupled phase model and its variants.  
These model are described by stochastic, nonlinear 
differential equations (DEs), which have been solved
by simulations or analytical methods such as the
Fokker-Planck equation (FPE) and the moment method.
In order to make our discussion concrete, we hereafter
pay our attention to the active rotator (AR) model,
which was first studied by 
Shinomoto and Kuramoto \cite{Shimokawa86}\cite{Sakaguchi88}. 
They obtained the FPE for 
the infinite-dimensional ($N=\infty$) AR model,
discussing the phase transition between the stationary state
and the time-periodic state.
Responses of excitable AR models to subthreshold inputs
have been investigated \cite{Wiesenfeld}-\cite{Tanabe01b}.
In recent years effects of additive and/or 
multiplicative noises in AR models have been studied by using 
FPE \cite{Park96}-\cite{Kim97b}.
Although FPE method is a powerful tool for solving
stochastic DEs, its application is limited to a single
or infinite system subject to white noises. 
Kurrer and Schulten proposed a moment method,
expanding FPE for the $N=\infty$ model
in a Taylor series around the center of 
distribution \cite{Kurrer95}.
Rodriguez and Tuckwell (RT) adopted a different moment
method 
in which the original stochastic DEs are expanded 
in a series of fluctuations around means 
of state variables \cite{Rod96}-\cite{Rod00}. 
RT's original moment
method was first applied to Fitzhugh-Nagumo (FN) neuron
model \cite{Rod96}\cite{Tanabe01} and 
then FN neuron networks \cite{Rod96}.
RT's moment method takes into account means, variances and covaricance
of local variables, then the number of DEs for  
$N$-unit FN neuron networks is $N_{eq}=N(2N+3)$, which is, for example, 
20300 for $N=100$. When we apply RT's moment method to
the $N$-unit AR network under consideration, we get $N_{eq}=(1/2)N(N+3)$,
which is 5150 for $N=100$. 
This exponential increase in the number
of DEs prevents us from calculations for a system
with a realistic size.

Quite recently the present author has proposed 
an alternative moment method, which is hereafter referred to as 
a dynamical mean-field approximation (DMA) \cite{Hasegawa02c}. In DMA
we take into account means, variances and covariances
of {\it local} and {\it global} variables, replacing the
original $2\:N$-dimensional DEs of the $N$-unit FN model
by eight-dimensional DEs independently of $N$.
We have investigated the $N$ dependence of
the spike timing precision, which has been shown to be improved
by increasing $N$.  
The purpose of the present paper is to apply DMA
to coupled AR networks to discuss their dynamics.
Although in our previous paper \cite{Hasegawa02c}, we included
up to the forth-order moments, we will, in this paper, take into account 
up to {\it infinite}-order moments, which are expressed as a product of 
second-order moments with the use of the Gaussian assumption
for the distribution of state variables.
The number of equations for $N$-unit AR networks becomes $N_{eq}=3$
in DMA, which is much smaller than $N_{eq}=(1/2)N(N+3)$ in the conventional
moment method.

The paper is organized as follows:
In Sec. II, we develop a DMA theory
for an ensemble of $N$ systems, 
obtaining equations of motions of moments
of local and global variables.
In Sec. III, the phase diagram showing various states 
in coupled AR models is discussed.
Discussions are given  
in Sec. IV, where we compare our DMA with the conventional 
moment method, showing that the former may
be alternatively derived from the latter with a proper
reduction of the number of variables with
the mean-field approximation.
The final Sec. V is devoted to conclusions.

%\newpage
\section{Dynamical mean-field approximation}

\subsection{Basic formulation}

We assume an ensemble consisting of coupled $N$-unit
phase models subject to white noises, whose
dynamics of the phase $\phi_i$ (mod $2\pi$) of the $i$th system
is described by nonlinear DEs given by 
\begin{eqnarray}
\frac{d \phi_{i}(t)}{dt} &=& F(\phi_i)
+ \frac{w}{N} \sum_j G(\phi_j-\phi_i)+ \xi_i(t),
\;\;\;\mbox{($i=1-N$)}
\end{eqnarray}
where explicit forms of $F(x)$ and $G(x)$ 
will be given shortly [Eq. (14)],
$w$ denotes the coupling, and
$\xi_i(t)$ is the independent
Gaussian white noise with $<\xi_i(t)>=0$ and
$<\xi_i(t)\:\xi_j(t')>=2D \; \delta_{ij}\:\delta(t-t')$,
the bracket $< \cdot >$ denoting the average
over stochastic random variables [see Eq. (4)].

We will express these nonlinear DEs
by moments of local and global variables
of the ensemble. 
The global variable is defined by
\begin{eqnarray}
\Phi(t)&=&\frac{1}{N}\;\sum_{i} \;\phi_{i}(t), 
\end{eqnarray}
and their averages by
\begin{eqnarray}
\mu(t)&=&<\Phi(t)>,
\end{eqnarray}
where the bracket $<>$ denotes
\begin{eqnarray}
<G(\mbox{\boldmath {$\phi$}})> &=& \int ...\int 
\;d \mbox{\boldmath {$\phi$}} 
\;G(\mbox{\boldmath {$\phi$}},t) \; 
p(\mbox{\boldmath {$\phi$}},t),
\end{eqnarray}
$p$({\boldmath $\phi$},$t$) being the probability 
distribution function (pdf)
for  the $N$-dimensional 
stochastic variable 
{\boldmath $\phi$}=$(\phi_1,....,\phi_N)$.

We express DEs given by 
Eq. (1) in terms
of the deviations from their average given by
\begin{eqnarray}
\delta \phi_i(t)&=& \phi_i(t)-\mu(t),
\end{eqnarray}
to get variances between local and global variables, given by 
(the argument $t$ is hereafter neglected)
\begin{eqnarray}
\gamma&=& \frac{1}{N}\; \sum_{i} <\delta \phi_i^2>, \\
\rho&=& <\delta \Phi^2>
=\frac{1}{N^2}\sum_i \; \sum_j <\delta \phi_i\; \delta \phi_j>, 
\end{eqnarray}
where $\delta \Phi = \Phi(t)-\mu(t)$.
It is noted that $\gamma$ expresses the
spatial average of fluctuations
in local variables of $\phi_i$
while $\rho$
denotes fluctuations in a global variable of $\Phi$.

We assume that the noise intensity $D$ is weak.
This allows us to
expand the right-hand side of Eq. (11) around the average $\mu$,
to get
\begin{eqnarray}
\frac{d \phi_i}{d t}&=& 
\sum_{\ell=0}^{\infty} \frac{F^{(\ell)}}{\ell !} 
\;(\delta \phi_i)^{\ell}
+ \frac{w}{N} \sum_j \sum_{\ell=0}^{\infty} 
\frac{G^{(\ell)}}{\ell !} (\delta \phi_j-\delta\phi_i)^{\ell}
+\xi_i(t), 
\end{eqnarray}
from which we get
\begin{eqnarray}
\frac{d \mu}{d t}&=& \frac{1}{N} \sum_i <\frac{d \phi_i}{dt}> \nonumber \\
&=&\frac{1}{N} \sum_i
\sum_{\ell=0}^{\infty} \frac{F^{(\ell)}}{\ell !} 
\;<(\delta \phi_i)^{\ell}>
+ \frac{w}{N^2} \sum_i \sum_j \sum_{\ell=0}^{\infty} 
\frac{G^{(\ell)}}{\ell !} <(\delta \phi_j-\delta\phi_i)^{\ell}>,
\end{eqnarray}
where $F^{(\ell)}=F^{(\ell)}(\mu)$ and 
$G^{(\ell)}=G^{(\ell)}(0)$.

We furthermore assume that pdf
of state variables takes Gaussian form.
Numerical simulations
have shown that for weak noises, the distribution of $\phi(t)$ 
in a single AR system nearly obeys
the Gaussian distribution,
although for strong noises, the distribution of $\phi(t)$
deviates from the Gaussian \cite{Tanabe01a}.
Similar behavior of the Gaussian distribution of state variables 
has been reported in FN and 
Hodgkin-Huxley neuron models \cite{Tanabe01a}\cite{Shimokawa99}.
When we adopt the Gaussian decoupling approximation,
averages higher than the second-order moments
in Eq. (9) may be expressed in terms of the
second moments given by
\begin{eqnarray}
<\delta \phi_1,...,\delta \phi_{\ell}>
&=&\sum_{all \;parings} \Pi_{km} <\delta \phi_k \delta \phi_m>, 
\hspace{1cm}  
\mbox{for even $\ell$}, \nonumber \\
&=& 0,\hspace{5cm}\mbox{for odd $\ell$},
\end{eqnarray}
where the summation is performed for
all $(\ell-1)(\ell-3)....3\cdot1$ combinations.
After some manipulations with the use of the
approximations mentioned above, we get equations of motions
for $\mu$, $\gamma$ and $\rho$ given by
(for details see Appendix A)
\begin{eqnarray}
\frac{d \mu}{d t}&=&\sum_{n=0}^{\infty} 
\frac{F^{(2n)}}{n!} (\frac{\gamma}{2})^n 
+ w\;\sum_{n=0}^{\infty} 
\frac{G^{(2n)}}{n!}(\gamma - \rho)^n, \\
\frac{d \gamma}{d t}&=& 2\;\gamma 
\sum_{n=0}^{\infty} \frac{F^{(2n+1)}}{n!} (\frac{\gamma}{2})^n 
+ 2 w (\gamma - \rho) \sum_{n=0}^{\infty} 
\frac{G^{(2n+1)}}{n!}(\gamma - \rho)^n
+ 2D, \\
\frac{d \rho}{d t}&=&2\;\rho
\sum_{n=0}^{\infty} \frac{F^{(2n+1)}}{n!} (\frac{\gamma}{2})^n 
%+ 2 w (\gamma - \rho) \sum_{n=0}^{\infty} 
%\frac{G^{(2n+1)}(\mu)}{n!}(\gamma - \rho)^n
+ \frac{2D}{N}. 
\end{eqnarray}

The coupled AR network is given by \cite{Shimokawa86}
\begin{eqnarray}
\frac{d \phi_{i}(t)}{dt} &=& c- a \;sin(x)
+ \frac{w}{N} \sum_j sin(\phi_j-\phi_i)+ \xi_i(t),
\;\;\;\mbox{($i=1-N$)}
\end{eqnarray}
with $F(x)=c- a \;sin(x)$ and $G(x)=sin(x)$ in Eq. (1).
In Eq. (14) 
$c \:(> 0)$ stands for the intrinsic frequency and $a$
the intensity of the pinning force introduced such that for $c < a$,
the system mimics the stochastic limit cycle 
or excitable elements.
The model with $c=0$ stands for the equilibrium planar model.
The case of $a=0$ corresponds to a usual phase model 
\cite{Kuramoto84}\cite{Strogatz00}.
For $w=0$ and $\xi_i=0$, the AR system
locates at the stationary point given by 
$\phi_i=\phi^*=arcsin(c/a)$.
When noises are introduced, the system
shows the intrigue behavior.  
Substituting $F(x)$ and $G(x)$ to Eqs. (11)-(13), we obtain
DEs for $\mu$, $\gamma$ and $\rho$ given by
\begin{eqnarray}
\frac{d \mu}{d t}&=&c - a \;sin(\mu) \;exp(-\frac{\gamma}{2}), \\
\frac{d \gamma}{d t}&=& -2 a \;\gamma \;cos(\mu)\;exp(-\frac{\gamma}{2}) 
- 2 w (\gamma - \rho)\; exp[-(\gamma-\rho)] + 2D, \\
\frac{d \rho}{d t}&=& -2 a\; \rho  \;cos(\mu)\;exp(-\frac{\gamma}{2})
+ \frac{2D}{N}, 
\end{eqnarray}
%On the contrary, when we adopt $G(x)=1-cos(x)$, Eqs.() become
%\begin{eqnarray}
%\frac{d \mu}{d t}&=&1 - a \;sin(\mu) \;exp(-\frac{\gamma}{2})
%+ w (1 -exp[-(\gamma-\rho)] ), \\
%
%\frac{d \gamma}{d t}&=& -2 a \;\gamma \;cos(\mu)\;exp(-\frac{\gamma}{2}) 
%+ \beta^2, \\
%
%\frac{d \rho}{d t}&=& -2 a \; \rho \;cos(\mu)\;exp(-\frac{\gamma}{2})
%+ \frac{1}{N}\beta^2, 
%\end{eqnarray}
The original $N$-dimensional {\it stochastic} DEs 
given by Eq. (1) are transformed to
three-dimensional {\it deterministic} DEs,
which show much variety depending on model parameters such
as $a$, $c$, $w$, $D$ and $N$.

We note that
the noise contribution is $2D$ in Eq. (12)
while that is $2D/N$ in Eq. (13).
It is easy to see that
\begin{eqnarray}
\rho&=&\gamma/N,
\hspace{2cm}\mbox{(for $w/D \rightarrow 0$)} \\
&=& \gamma.
\hspace{2cm}\mbox{(for $D/w \rightarrow 0$)}
\end{eqnarray}
Equation (18) agrees with the {\it central-limit theorem}.
%, and
%which is the origin of the pooling effect \cite{deCharms00}.
In the limit of $N=\infty$, we get $\rho=0$.
On the contrary,
in the limit of $N=1$, we have $\rho=\gamma$.

\subsection{Various quantities}

\vspace{0.5cm}
\noindent
{\bf Distribution of local variables}

Adopting the mean-field approximation,
we get $<\phi_i> \simeq (1/N) \sum_k <\phi_k>=\mu$ and 
$<\delta \phi_i^2> \simeq (1/N) \sum_k <\delta \phi_k^2> = \gamma$.
Then the distribution for the
variable $\phi_i$ is given by
\begin{eqnarray}
P(\phi_i) 
%&=& \int ... \int \; \Pi_{j(\neq i)} \:dx_j \:\Pi_j \:dy_j\;
%p(x_1, ...,x_N, y_1, ...,y_N), 
&\simeq&(\frac{1}{\sqrt{\gamma}})
\;\overline{\phi}(\frac{\phi_i-\mu}{\sqrt{\gamma}}),
\end{eqnarray} 
where $\overline{\phi}(x)$ is the normal distribution function given by
\begin{equation}
\overline{\phi}(x)=\frac{1}{\sqrt{2 \pi}} 
{\rm exp}(-\frac{x^2}{2}).
\end{equation} 
%with
%\begin{equation}
%\sigma_{\ell}=\sqrt{\gamma}.
%\end{equation}
The probability given by Eq. (20) depends on the time because
of the time dependence of $\mu(t)$ and $\gamma(t)$.

\vspace{0.5cm}
\noindent
{\bf Distribution of global variables}

Mean and variance of global variables $\Phi$
are given by $<\Phi>=\mu$ and 
$<\delta \Phi^2>=\rho$, respectively.
We get the distribution  for the
global variable $\Phi$ given by
\begin{eqnarray}
P(\Phi) 
%&=& \int dY  \; p(X,Y), \\
&\simeq& (\frac{1}{\sqrt{\rho}})
\; \overline{\phi}(\frac{\Phi-\mu}{\sqrt{\rho}}).
\end{eqnarray} 
%with
%\begin{equation}
%\sigma_g=\sqrt{\rho}.
%\end{equation} 

\vspace{0.5cm}
\noindent
{\bf Averaged frequency}

The averaged frequency $\nu$ is defined by
\begin{equation}
\nu= [\frac{1}{N (N_{fi}-1)}\; \sum_i \sum_k T_{oik}]^{-1},
\end{equation}
with
\begin{equation}
T_{oik}=t_{i k+1}-t_{i k},
\end{equation}
\begin{equation}
t_{ik}=\{t \mid \phi_i(t)=\theta; \dot{\phi_i} > 0; 
t \geq t_{ik-1}+\tau_r \},
\end{equation}
where $N_{fi}$ stands for the number of firings
of a given rotator $i$, $\dot{\phi_i}=d \phi_i/dt$,
$T_{oik}$ the interspike interval (ISI) of output signals,
$t_{i k}$ the $k$th firing time, and
$\theta \;(=2\:\pi)$ and $\tau_r\;(=5)$ are the threshold level
and the refractory period, respectively.
When there is no firings, we set $\nu=0$.

\vspace{0.5cm}
\noindent
{\bf Order parameters }

The order parameter $\zeta$ and 
its fluctuation $\delta \zeta$ are defined by
\begin{eqnarray}
\zeta&=&\overline{\mid z(t) \mid}, \\
\delta \zeta &=& \sqrt{\overline{\mid z(t) \mid^2}-\zeta^2},
\end{eqnarray}
with
\begin{eqnarray}
z(t) &=& \sum_i exp[i\; \phi_i(t)]  
= <exp[i\; \phi_i(t)]>,
\end{eqnarray}
where the overline denotes the temporal avarage.
By expanding $z(t)$ in a series of $\delta \phi_i$
around $\mu$ and
adopting the Gaussian decoupling approximation given by Eq. (10),
we get
\begin{equation}
z(t) = exp[i\; \mu(t)]\;\sum_n \frac{1}{n!}(\frac{-\gamma}{2})^n
=exp[i\; \mu(t) - \frac{\gamma(t)}{2}]
\end{equation}

\vspace{0.5cm}
\noindent
{\bf Synchronization ratio}

The synchronization ratio $\sigma$
is defined by 
\begin{equation}
\sigma=\overline{s(t)},
\end{equation}
with
\begin{equation}
s(t)=\frac{(\rho/\gamma-1/N)}{(1-1/N)}.
\end{equation}
For the completely synchronous (asynchronous) state,
both $\zeta$ and $\sigma$ become 1 (0). It is noted, however, that
while $\zeta$ depends on $\gamma$,
$\sigma$ is a function of 
$(\rho/\gamma-1/N) = N^{-1}\;\sum_{i \neq j} 
<\delta \phi_i\:\delta \phi_j>/
\sum_i \; <\delta \phi_i^2>$; the ratio of the inter-AR correlation
to the intra-AR correlation.

\vspace{0.5cm}

Before discussing calculated results with the use of DMA,
it is worth to mention the calculation 
of Kurrer and Schulten \cite{Kurrer95}.
They intended to solve the FPE 
for the $N=\infty$ AR model
given by \cite{Shimokawa86}
\begin{equation}
\frac{\partial}{\partial t} \;n(\phi,t)
= - \frac{\partial}{\partial \phi}
[F(\phi,t) - w \int d \phi' \; sin(\phi-\phi') n(\phi',t)] \;n(\phi,t)
+D \frac{\partial^2}{\partial t^2}\;n(\phi,t),
\end{equation}
where the density probability $n(\phi, t)$ is defined by
\begin{eqnarray}
n(\phi, t)&=& \frac{1}{N} \sum_i \delta(\phi-\phi_i(t)), 
\end{eqnarray}
with the periodic condition: $n(\phi+2 \pi,t)=n(\phi,t)$
and the normalization condition: 
$\int \;d \phi \;n(\phi,t)=1$.
Kurrer and Schulten \cite{Kurrer95}
expanded $F(\phi, t)$ in a Taylor series around the center of 
distribution, assuming the Gaussian form for $n(\phi, t)$ given by
\begin{eqnarray}
n(\phi, t)&=&\frac{1}{\sqrt{2 \pi\:v(t)}}\;
exp \left( -\frac{[\phi-u(t)]^2}{2 \;v(t)} \right),
\end{eqnarray}
where
the mean $u(t)$ and variance $v(t)$ obey DEs given by
\begin{eqnarray}
\frac{d u}{d t}&=&c - a \;sin(u) \;exp(-\frac{v}{2}), \\
\frac{d v}{d t}&=& -2 a \;v \;cos(u)\;exp(-\frac{v}{2}) 
- 2 w \;v\; exp(-v) + 2D. 
\end{eqnarray}
Equations (35) and (36) resemble our Eqs. (15)-(17)
if we read $u \rightarrow \mu$ and $v \rightarrow \gamma$. 
Actually,
Eqs. (35) and (36) are equvalent to Eqs. (15) and (16)
in the case of $\rho=0$, which is realized
in the limit of $N=\infty$.

%\newpage
\section{Calculated results}

DMA equations given by Eqs. (15)-(17) have been solved
by the forth-order Runge-Kutta method with a time step of 0.01,
the initial conditions being $\mu(0)=\gamma(0)=\rho(0)=0$.
Calculations have been performed for $0 \leq t \leq 1000$ (100 000 steps) and
results for $t < 100$ are discarded.
Simulations of directly solving Eq. (1) have been made by 
the forth-order Runge-Kutta method with a time step of 0.01,
the initial conditions being $\phi_i(0)=0$ ($i=1$ to $N$).
The number of trials for a given set of parameters
in our simulations is hundred
otherwise noticed. We have solved also
FPE given by Eq. (32), which is valid 
for the $N=\infty$ AR model. We first 
Fourier transform FPE
with the first 30 modes after Ref. \cite{Shimokawa86}. 
A set of 61 ordinary DEs has been solved by the Runge-Kutta method.

\subsection{Phase diagram for various types of solutions}

By solving Eqs. (15)-(17),
we get the stationary state and
the non-stationary state:
in the former state the variables are time
independent while in the latter state they 
are time dependent. 
The equilibrium values of $\mu$, 
$\gamma$ and $\rho$ in the stationary state
are given by (we set $c=1$ hereafter)
\begin{eqnarray}
\mu&=&arcsin[(\frac{1}{a}) \;exp(\frac{\gamma}{2})], \\
\gamma&=& \frac{D +w \;\rho \;exp[-(\gamma-\rho)]}
{\sqrt{a^2 \;exp(-\gamma)-1}+w \;exp[-(\gamma-\rho)]}, \\
\rho&=& \frac{D/N}{\sqrt{a^2 \;exp(-\gamma)-1}}.
\end{eqnarray}
The stationary state where 
Eqs. (37)-(39) are satisfied, is hereafter referred to as the S state.
Kurrer and Schulten \cite{Kurrer95} pointed out that
the non-stationary state may be classified into the time periodic
(P) state and the random, disordered (R) state.
DMA also yields three types of the S, P and R states
characterized by the quantities
of $\zeta$, $\delta \zeta$, $\nu$ and $\sigma$ introduced in Sec. IIB,  
result being summarized in Table 1.
In the S state, $\delta \zeta$ and $\nu$ are vanishing 
while $\zeta \;(\simeq 1)$
and $\sigma$ are finite.
In the P state, all quantities are finite.
In contrast, in the R state, all quantities except $\nu$ vanish.
%We will shortly discuss
%the P' state which is characterized by non-vanishing $\nu$ and
%$\sigma$ with vanishing $\zeta = \delta \zeta=0$.

Boundaries between these three states depend 
on $a$, $w$, $D$ and $N$. 
Figure 1 expresses the $D-a$ phase diagram showing the boundaries 
between these states in coupled AR models calculated
with the use of DMA for $N=10$, 100 and $N=\infty$.
The gradient of the boundary between the stationary (S) state and
non-stationary (P+R) states is decreased as increasing the value of $w$
and/or of $N$.
The difference between boundaries for $N=10$ and $N=\infty$ with $w=0.1$ is
very small:
the effect of $N$ becomes more significant for a stronger coupling. 
The critical $a$ value, $a_c$, above which the S state exists,
is given by
\begin{equation}
a_c - 1 \simeq [c_1 - c_2\; w\; (1 - \frac{1}{N})]\;D,
\end{equation}
where $c_1=2.25$ and $c_2=1.75$.
In contrast, the critical value of $a_d$ for the 
boundary between the P and R states for  $w=1.0$
is given by
\begin{equation}
a_d - 1 \simeq - \;(d_1+\frac{d_2}{N} )[D - (d_3 + \frac{d_4}{N})],
\end{equation}
where $d_1=5.36$, $d_2=257$, $d_3= 0.265$ and $d_4=0.9$.

The behavior of the solutions of DMA in the
S, P and R states
when $D$ and/or $a$ values are changed, is shown in 
Figs. 2-5.
We will first mention the calculations of DMA in the three states.
%and then compare them with results of simulations 
%and the FPE method.
Figure 2(a) and 2(b) express the $D$ dependence of $\zeta$,
$\delta \zeta$, $\nu$ and $\sigma$  
for a representative set of parameters of 
$a=1.05$, $w=1.0$ and $N=100$,
showing that
the network is in the S state
for $D \leq 0.082$, in the P state for $0.082 < D \leq 0.273$,
and in the R state for $D > 0.273$.
In contrast, Fig. 3(a) and 3(b) express $\zeta$,
$\delta \zeta$, $\nu$ and $\sigma$ as a function of $a$
for a set of parameters of 
$D=0.1$, $w=1.0$ and $N=100$,
for which the networks is in the S state for $a \geq 1.06$
and in the P state for $a < 1.06$.

Equations (37)-(39) for small $D$ and $w$ in the S state yield 
\begin{eqnarray}
\mu &=& arcsin(\frac{1}{a})+d_1\; D + .., \\
\gamma &=& \frac{D}{\sqrt{a^2-1}}
[1 + d_2 \;D - d_3 (1-\frac{1}{N}) \;w + d_4 w^2+ ..], \\
\rho &=& \frac{(D/N)}{\sqrt{a^2-1}}[1+d_2 \;D+ ..,], 
\end{eqnarray}
leading to
\begin{eqnarray}
\zeta &=& 1 - \frac{\gamma}{2}, \\
\sigma &=&\frac{1}{N}\; \frac{d_3 \:w}{(1+d_2 \:D)},
\end{eqnarray}
where $d_1=1/2(a^2-1)$, $d_2=a^2/2 (a^2-1)^{3/2}$,
$d_3=1/\sqrt{a^2-1}$, and $d_4 \;(>\; 0)$ is a complex
function of $D$, $w$ and $N$.

Solid curves in Figs. 4(a) and 4(b) show distributions of 
local [$P(\phi_i(t))$] and global variables [$P(\Phi(t))$], 
respectively, in DMA
for $a=1.05$, $w=1.0$, $D=0.05$.
They are obtained by
Eqs. (20) and (22) with
$\mu=1.339$, $\gamma=0.04354$
and $\rho=0.00212$.

Figures 2(a) and 2(b) show that
when the noise intensity is increased 
and crosses the value of 0.082, the AR network begins 
correlated firings. This implies the appearance of 
the P state, where 
$\delta \zeta$, $\nu$ and $\sigma$ are
continuously changed.
Solid curves in Fig. 4(c) ad 4(d) express $P(\phi_i(t))$
and $P(\Phi(t))$ for $D=0.10$, respectively, which are given
by Eqs. (20) and (22) with $\mu=1.497$, $\gamma=0.11022$
and $\rho=0.009443$.
The time evolution of the probability of $P(\phi(t))$ 
calculated in DMA 
for $D=0.10$ in the P state
is shown in Fig. 5(a), 
which is oscillating with the period of about 40.
It is noted that not only the position of $P(\phi(t))$
but also its width change as a function of $t$.
For example, we get $\mu=6.151$ and $\gamma=1.711$ at $t=120$
while $\mu=1.497$ and $\gamma=0.11022$ at $t=100$.

When the $D$ value is rmore increased,
the AR network fires abundantly 
and irregularly, which suggests the appearance of the R state.
The solution of Eqs. (37)-(39) in the R state for a large $t$
is given by
\begin{eqnarray}
\mu &\simeq& c \: t, \\
\gamma &\simeq& 2 D t, \\
\rho &\simeq& (\frac{2D}{N}) \;t,
\end{eqnarray}
which lead to vanishing $\zeta$, $\delta\zeta$ and $\sigma$ except
$\nu$.
Figure 2(a) and 2(b) show that 
$\zeta$ and $\delta \zeta$ suddenly vanish
at $D = 0.273$ with no hysteresis.
Solid curves in Fig. 4(e) ad 4(f) express $P(\phi_i(t))$
and $P(\Phi(t))$ for $D=0.30$, respectively, which are given
by Eqs. (20) and (22) with $\mu=1.339$,
$\gamma=19.9$ and $\rho=0.199$.

In the following, results of DMA will be compared
with those of simulations and FPE.
Dashed curves in Figs. 2(a) and 2(b)
show the results of simulations 
for the $D$ dependence of $\zeta$, $\delta \zeta$, 
$\nu$ and $\sigma$.
The agreement of $\zeta$ in DMA with that in simulations
is good for S and P states. However, it is not good in the R state,
where $\zeta$ vanishes in DMA but not in simulations. 
This is expected to be due to deviations of the state-variable 
distributions from the Gaussian form.
When $D$ is more increased in the R state, our simulations yields
a gradual decrease in $\zeta$,
which is 0.454, 0.276, 0.188, 0.139 and 0.110 for $D=1.0$,
2.0, 3.0, 4.0 and 5.0, respectively,
with $a=1.05$, $w=1.0$ and $N=100$.
%a lack of trial numbers of
%$N_{tr}=100$ in our simulations, $N_{tr}=\infty$ being implicitly assumed
%in deriving DMA.
We note in Fig. 2(a) that $\delta \zeta$ of simulations is
about ten times smaller than that of DMA.
This is clearly seen in Fig. 6(a) where we plot the time evolution
of $\mid z(t) \mid$ obtained by DMA and simulations
for $a=1.05$, $w=1.0$, $D=0.10$ and $N=100$.
The former has larger temporal fluctuations than the latter
although both yield similar averaged values of $\zeta = \overline{\mid z(t) \mid}$. 
In contrast, Fig. 6(b) shows the time dependence of $s(t)$ calculated
by DMA and simulations for $a=1.05$, $w=1.0$, $D=0.10$ and $N=100$.
Again our DMA yields larger
fluctuations in $s(t)$ than simulations although 
both methods lead to similar averaged values of $\sigma = \overline{s(t)}$.
When comparing K(a) with K(b), we notice that 
the time dependence of $\mid z(t) \mid$ is not the same as that of $s(t)$.
%when $z(t)$ is increased $s(t)$ is decreased, and vice versa. 
This is because $z(t)$ is a function of $\gamma$ [Eq. (29)] 
while $s(t)$ is a function of the ratio of $\rho/\gamma$ [Eq. (31)].
Dashed curves in Figs. 3(a) and 3(b)
show the $a$ dependence of $\zeta$, $\delta \zeta$, 
$\nu$ and $\sigma$ obtained by simulations. 
$P(\phi)$ and $P(\Phi)$ obtained by simulations are plotted
by dashed curves in Fig. 4(a)-4(f).
Dotted curves in Fig. 4(a), 4(c) and 4(e) denote
$n(\phi)$ obtained by FPE for 
the $N=\infty$ AR model.
Figure 5(b) express the time evolution of
$n(\phi, t)$ in the P state obtained by FPE.
From a comparison between the results
of DMA and simulations (and FPE) mentioned above, we note
that DMA is good for the S state, in fairly good for
the P state in the qualitative sense,
but not good for the R state.

For a comparison, we show by
the dotted curve in Fig. 1, the boundary obtained 
by Shimokawa and Kuramoto (SK)
with the use of the FPE for $w=1.0$ and $N=\infty$
\cite{Shimokawa86}. 
The ordered P state where $\delta \zeta \neq 0$ 
and $\nu \neq 0$ is reported to exist in
the triangle region enclosed by the dotted curve and the horizontal axis.
The P state obtained by SK is nearly in agreement with our P state.
In SK, states besides the P state are regarded 
as the stationary state where $\partial n(\phi,t)/\partial t = 0
$\cite{Shimokawa86}.
On the contrary,
Kurrer and Schulten (KS)
distinguished the R state from the P state, both
of which are non-stationary ($\nu \neq 0$) \cite{Kurrer95}. 
The results of SK and KS are for the $N=\infty$ AR model.
Figures 7(a) and 7(b) show the $D$ dependence 
of $\zeta$, $\delta\zeta$ and $\nu$
for a set of parameters of $a=1.05$, $w=1.0$ and $N=10^4$,
which are the same as
in Figs. 5(a) and 5(b) except $N$.
In order to simulate the $N=\infty$ limit, the $N$ value  
in Figs. 7(a) and 7(b) is chosen to be very
large but finite
because $s(t)$ given by Eq. (31) is not properly defined
in this limit.
Results for $N=\infty$ in Figs. 1 and 7 
should be compared with those obtained by SK and KS. 
As was pointed in Sec. IIB,
DEs of DMA given by Eqs. (15)-(17) in the limit of $N=\infty$ agree with
those of KS given by Eqs. (35)-(36). Nethertheless,
our $D-a$ phase diagram for $N=\infty$ in Fig. 1 does not agree
with that of KS. For example, KS obtained the critical
values given by \cite{Kurrer95}
\begin{eqnarray}   
a_c - 1 &=& \frac{D}{2\;w}, \hspace{1.5cm}\mbox{(between S and P + R)} \\
\frac{D_d}{w} &=& 0.736, \hspace{1.5cm}\mbox{(between P and R)}
\end{eqnarray}
which do not agree with our expressions given 
by Eqs. (40) and (41) for $N=\infty$.

\subsection{Cluster-size ($N$) dependence in the S state}
%\vspace{0.5cm}

Since one of the advantages of DMA is that 
we can discuss the finite-$N$ property
of coupled AR networks, 
we have made more detailed calculations of the
$N$ dependence of the quantities in the S state.
Figures 8(a) and 8(b) show the log-log plot of
the $N$ dependence of 
$\gamma$, $\rho$ and $\sigma$,
results for $N=10$ and 20 being for 500 trials.
Solid curves in Fig. 8(a) express the result of DMA
for a set of parameters of $a=1.05$, $w=1.0$ and $D=0.05$,
whereas circles, squares and triangles 
denote those of simulations. For this set of
parameters, the P state is realized for $N \leq 9$.
We note that as $N$ is decreased from above and approaches to the S-P boundary,
fluctuations of $\gamma$ and $\rho$ are increased
(and $\sigma$ is also increased).
Similar behavior is observed for a different set of
parameters. Figure 8(b) shows the results
for $a=1.20$, $w=1.0$ and $D=0.10$.
With this set of parameters, we get the S state for
$N \geq 2$ and the P state for $N=1$.
Figures 8(a) and 8(b) show that as increasing $N$, $\rho$ is 
much decreased but $\gamma$ shows only a weak $N$ dependence. 
We note that $\sigma$ is decreased as increaing $N$
whereas $\zeta=exp(-\gamma/2)$ shows little $N$ dependence.
We should stress that although
fluctuations of global variables is 
inversely decreased as $\rho \propto 1/N$ consistent with
the central-limit theorem,
those of local variables remain finite 
even for $N \rightarrow \infty$.

%\newpage
\section{Discussions}

We have proposed DMA theory for stochastic, nonlinear networks
like coupled AR models, 
taking into account means, variances and covariances of {\it local}
and {\it global} variables.
It is worth to compare DMA with the conventional
moment method in which means, variances and covariances 
of local variables are given by
\begin{eqnarray}
m_i &=& <\phi_i>, \\
C_{ij}&=& <\Delta \phi_i \;\Delta \phi_j>,
\end{eqnarray}
with $\Delta \phi_i = \phi_i-m_i$.
By using the Gaussian decoupling approximation [Eq. (10)],
we get (for details see Appendix B)
\begin{eqnarray}
\frac{d m_i}{d t}&=&\sum_{n=0}^{\infty} 
\frac{F_i^{(2n)}}{2^n\;n!} C_{ii}^n
+ \frac{w}{N} \sum_k \sum_{n=0}^{\infty} 
\frac{G^{(2n)}}{2^n\;n!} [C_{kk}+C_{ii}-2 C_{ik}]^n,\\
\frac{d C_{ij}}{dt}&=& \sum_{n=0}^{\infty}
\frac{F_i^{(2n+1)}}{2^n\;n!}(C_{jj}^n +C_{ii}^n)\;C_{ij} 
+\frac{w}{N} \sum_k \sum_{n=0}^{\infty}
\frac{G^{(2n+1)}}{2^n\;n!}  \nonumber\\
&&
\times[(C_{kk}+C_{ii}-2 C_{ik})^n\;(C_{jk}-C_{ij})
+(C_{kk}+C_{jj}-2 C_{jk})^n\;(C_{ik}-C_{ij})]
+ 2 D_i \delta_{ij},
\end{eqnarray}
where $F_i^{(\ell)}=F^{(\ell)}(m_i)$ and 
$G^{(\ell)}=G^{(\ell)}(0)$.

For the AR model, Eqs. (54) and (55) become
\begin{eqnarray}
\frac{d m_i}{dt}&=& c -a \;sin(m_i) \;exp(-\frac{1}{2} C_{ii}), \\
\frac{d C_{ij}}{dt} &=& -a \;cos(m_i)
\sum_{n=0}^{\infty} \frac{(-1)^n}{2^n \; n!} (C_{ii}^n+C_{jj}^n) C_{ij} 
+ 2 D \delta_{ij}  \nonumber\\
&&+ \frac{w}{N} \sum_k \sum_{n=0}^{\infty}
\frac{(-1)^n}{2^n\;n!}
[(C_{ii}+C_{kk}-2 C_{ik})^n(C_{jk}-C_{ij})  \nonumber\\
&&+(C_{jj}+C_{kk}-2 C_{jk})^n(C_{ik}-C_{ij})] ,
\end{eqnarray}
For variances ($i=j$), Eq. (57) becomes
\begin{eqnarray}
\frac{d C_{ii}}{dt} &=& -2 a \;cos(m_i) \;C_{ii} 
\;exp(-\frac{1}{2} C_{ii})
+ 2 D  \nonumber\\
&&+ \frac{2w}{N} \sum_k \sum_{n=0}^{\infty}
\frac{(-1)^n}{2^n\;n!}
[(C_{ii}+C_{kk}-2 C_{ik})^n(C_{ik}-C_{ii})].
\end{eqnarray}

Taking into the symmetry relations: 
$C_{ij}=C_{ji}$, we get 
the number of DEs in the moment method
to be $N_{eq}\;= N(N+3)/2$, which is
65, 5150 and 501 500
for $N=$10, 100 and 1000, respectively, 
while $N_{eq}=3$ in our DMA.

It will be shown that
we can derive DMA from the moment method
by reducing the number of DEs,
adopting the mean-field approximation:
\begin{eqnarray}
m_i&\simeq&\mu, \\
C_{ii}^n &\simeq& \gamma^{n-1} C_{ii}, \\
(C_{kk}+C_{ii}-2C_{ik})^n 
&\simeq& 2^{n-1}(\gamma-\rho)^{n-1} 
(C_{kk}+C_{ii}-2C_{ik}),
\hspace{1cm}\mbox{($i \neq k$)}
\end{eqnarray}
with the relations given by
\begin{eqnarray}
\mu=\frac{1}{N} \sum_i m_i, \\
\gamma = \frac{1}{N} \sum_i C_{ii}, \\
\rho= \frac{1}{N^2} \sum_i \sum_j C_{ij}. 
\end{eqnarray}
DEs for $\mu$, $\gamma$ and $\rho$
are given by Eqs.(11)-(13) for the general phase model
or by Eqs. (15)-(17) for the AR model.

It is possible to regard DMA as the 
{\it single-site self-consistent} theory.
Let us assume a configuration in which a {\it single}
nonlinear system $i$ is embedded 
in an effective medium whose effect
is realized by a given system $i$
as its effective external input
through the coupling $w$.
Assuming $m_i=\mu$,
we replace quantities 
in coupling terms of Eqs. (57) and (58) 
by effective quantities of $\mu$, $\gamma$ and $\rho$,
to get
\begin{eqnarray}
\frac{d m_i}{d t}&=&\sum_{n=0}^{\infty} 
\frac{F^{(2n)}}{2^n\;n!} C_{ii}^n
+ \frac{w}{N} \sum_k \sum_{n=0}^{\infty} 
\frac{G^{(2n)}}{n!} (\gamma-\rho)^{n}, \\
\frac{d C_{ij}}{dt}&=& \sum_{n=0}^{\infty}
\frac{F^{(2n+1)}}{2^{n}\;n!}\;(C_{jj}^n+C_{ii}^n)^n\;C_{ij} 
+\frac{w}{N} \sum_k \sum_{n=0}^{\infty}
\frac{G^{(2n+1)}}{n!} (\gamma-\rho)^{n+1}.
\end{eqnarray}
We should note that $m_i$ and $C_{ij}$ determined by Eqs. (65) 
and (66) are functions of $\mu$, $\gamma$ and $\rho$.
In order to self-consistently determine them,
we impose the self-consistent conditions given by
\begin{eqnarray}
\mu&=& m_i, \\
%(=\frac{1}{N} \sum_i m_{\kappa}^i), \\
\gamma_&=& C_{ii}, \\
%(= \frac{1}{N} \sum_i C_{\kappa,\lambda}^{i,i}), \\
\rho 
&=& \frac{1}{N} \sum_{j} C_{ij}.
%(= \frac{1}{N^2} \sum_i \sum_{j (\neq i)} C_{\kappa,\lambda}^{i,j}).
\end{eqnarray}
Note that Eqs. (67)-(69) are assumed to hold independently of $i$ and
that $m_i$
and $C_{ij}$ in their right-hand sides 
are functions of
$\gamma$
and $\rho$.
The condition given by Eqs. (65)-(69) 
with the mean-field approximation given by Eq. (59)-(61)
yields DEs for
$\gamma$
and $\rho$
which are again given by Eqs. (11)-(13).
The self-consistent condition given by Eq. (67)-(69), which 
assumes that the quantities averaged at a given site are the same as
those of the effective medium, is common in mean--field theories
such as the Weiss theory for magnetism \cite{Weiss07}
%BCS theory for superconductivity 
and the coherent-potential 
approximation for random alloys \cite{Soven67}.

By using DMA,
we have investigated the response of the 
excitable, coupled AR networks
to an applied periodic spike, by adding
to the right-hand side of Eq. (15), the input term
given by
\begin{eqnarray}
I_{in}(t)&=& g, \hspace{1cm} \mbox{for $m\:T_p \leq t 
< m\:T_p + T_w$ (m: integer)} \nonumber \\
&=& 0, \hspace{1cm} \mbox{otherwise}
\end{eqnarray}
where $g$ denotes the magnitude, 
and $T_p$ (=50) and $T_w$ (=5) stand for
the period and the duration of spikes, respectively.
We get the critical value of $g_c=0.159$ below which 
there are no firings for $D=0$.
Figure 9(a) shows the distribution of ISI, $T_{o}$, 
of output signals 
defined by Eqs. (24) and (25) as a function of $D$
in the absence of input spikes ($g=0$) for $a=1.05$, $w=1.0$ and
$N=100$. Firings begin at $D=0.082$, above which the system is in the P state
as discussed in Sec. IIIA. Around the P-R transition at $D=0.273$, ISIs have
a small distribution.
When the input spike is applied, distributions of ISIs are 
significantly changed.
Figure 9(b) shows the distribution when the subthreshold input with
$g=0.1 $ $(< g_c)$ is applied.
Firings occur at $D \geq 0.04$ with a help of noises.
A flat segment at $0.04 < D < 0.08$ corresponds to a periodic solution
locked to input spikes while the others show the complex behavior.
In contrast, Fig. 9(c) shows the distribution of ISIs for
the suprathreshold input with $g=0.2$.
At $0.05 < D < 0.08$ in the S state, 
a new branch with $35 < T_o < 43$ appears beside
the branch with $T_o = 50$ locked to inputs.
The distribution of ISIs in the presence of input spikes
has much variety than that in the absence of noises, in particular
in the P state, where the bifurcation is realized as 
the noise intensity is changed.

It is possible to discuss the firing-time accuracy of
excitable AR models for an external input
with the use of DMA \cite{Hasegawa02c}.
%For a simplicity, a single spike is assumed to be applied 
%to the AR network at $t=t_{in}$.
The $k$th firng time of a given rotator $i$ is defined as the
time when $\phi_i(t)$ crosses the threshold $\theta$ 
from below [Eq. (25)]:
\begin{equation}
t_{ik}=\{t \mid \phi_i(t)=\theta; \dot{\phi_i} > 0; 
t \geq t_{ik-1}+\tau_r \}.
\end{equation}
By using the discussion presented in Sec. IIB, the probability
$W_{\ell}$ when $\phi_i(t)$ at $t$ is above the threshold $\theta$
is given by
\begin{equation}
W_{\ell}(t)= 1 - \psi(\frac{\theta-\mu}{\sqrt{\gamma}}),
\end{equation}
where $\psi(y)$ is the error function given by integrating
the normal distribution function $\overline{\phi}(x)$
from $-\infty$ to $y$ [Eq. (21)].
The fraction of a given rotator $i$ emitting output at $t$
is given by
\begin{equation}
Z_{\ell}(t)=\frac{dW_{\ell}}{dt} \:\Theta(\dot{\mu})
=\overline{\phi}(\frac{\theta-\mu}{\sqrt{\gamma}})
\frac{d}{dt}(\frac{\mu}{\sqrt{\gamma}})\Theta(\dot{\mu}),
\end{equation}
where $\Theta(x)=1$ for $x\geq 0$ and 0 otherwise,
and $\dot{\mu}=d\mu(t)/dt$.
When we expand $\mu(t)$ in Eq. (73) around $t_o^*$ where
$\mu(t_o^*)=\theta$, we get 
\begin{equation}
Z_{\ell}(t) \sim \overline{\phi}(\frac{t-t_o^*}{\delta t_{o\ell}})
\frac{d}{dt}(\frac{\mu}{\sqrt{\gamma}})\Theta(\dot{\mu}),
\end{equation}
with 
\begin{equation}
\delta t_{o\ell}=\frac{\sqrt{\gamma}}{\dot{\mu}}.
\end{equation}
We note that $Z_{\ell}$ provides the distribution
of firing times, showing that most of firings locate 
in the range given by
\begin{equation}
t_{o\ell} \in [t_o^* - \delta t_{o\ell}, \;t_o^* + \delta t_{o\ell}].
\end{equation}
In the limit of vanishing $D$, Eq. (74) reduces to
\begin{equation}
Z_{\ell}(t)=\delta(t-t_o^*).
\end{equation}

Similarly, we define the $k$th firing time relevant to 
the global variable $\Phi(t)$ as
\begin{equation}
t_{gk}=\{t \mid \Phi_i(t)=\theta; \dot{\Phi_i} > 0; 
t \geq t_{gk-1}+\tau_r \}.
\end{equation}
The distribution of firing times $t_{g}$ is given by
\begin{equation}
Z_{g}(t) \sim \overline{\phi}(\frac{t-t_o^*}{\delta t_{og}})
\frac{d}{dt}(\frac{\mu}{\sqrt{\rho}})\Theta(\dot{\mu}),
\end{equation}
with 
\begin{equation}
\delta t_{og}=\frac{\sqrt{\rho}}{\dot{\mu}}.
\end{equation}
Equation (79) shows that most of $t_{og}$ locate in the range given by
\begin{equation}
t_{og} \in [t_o^* - \delta t_{og}, \;t_o^* + \delta t_{og}].
\end{equation}
From Eqs. (75) and (80), we get
\begin{equation}
\frac{t_{og}}{t_{o\ell}}=\sqrt{\frac{\rho}{\gamma}}
\rightarrow \frac{1}{\sqrt{N}}. \hspace{1cm} 
\mbox{(as $w/D \rightarrow 0$)}
\end{equation}
This implies that the firing-time accuracy is improved
as the ensemble size is increased even when there no couplings
among ARs. This is consistent with results reported
prevously \cite{Pei96a}-\cite{Hasegawa02b}.

\section{Conclusions}

%We have proposed DMA theory for stochastic, nonlinear networks
%like coupled AR models, 
%taking into account means, variances and covariances of {\it local}
%and {\it global} variables.
We have developed DMA,
which has been shown to be derived in various ways:
equations of motions for means, variances and covariances
of local and global variables (Sec. IIA), a reduction in
the number of moments in the moment method, and 
a single-site self-consistent approximation to 
the moment method (Sec. IV).
Our DMA theory, which assumes weak noises 
and the Gaussian distribution of state variables,
goes beyond the weak coupling because
no constraints are imposed on the coupling strength.
The advantage of DMA is that it can be applied to large but
finite-$N$ nonlinear systems subject 
not only to white noises but also to color noises.
This is in contrast with FPE, which is applicable
to a single or infinite system subject to white noises.
The limitation of DMA is the weak noise, for which
calculated results based on DMA are in fairly good agreement
with those obtained by direct simulations.
When the noise intensity becomes stronger,
the state-variable distribution more deviates from the 
Gaussian form, 
and the agreement of results
of DMA with those of simulations becomes worse.
Nevertheless, our DMA is expected to be meaningful
for qualitative or semi-quantitative discussion
on the properties of coupled nonlinear systems.
It is possible to regard DEs given by Eqs. (15)-(17)
as the mean-filed AR model which may show interesting
behavior for applied input signals and noises.
When we consider an ensemble
of $N$-unit systems, each of which is described by 
a $M$-variable nonlinear DE,
the number of the deterministic DEs is
$N_{eq}=M+M(M+1)=M(M+2)$ independently of $N$ in DMA 
while it is $N_{eq}=NM+(1/2)NM(NM+1)=(1/2)NM(NM+3)$ 
in the conventional moment. 
In the case of $M=2$ (as in FN model), for example,
DMA leads to $N_{eq}=8$
while the moment method yields
$N_{eq}=N(2N+3)$, which is 2310, 20 300 and
2 003 000 for $N=10$, 100 and 1000, respectively. 
These figures clearly show the advantage and feasibility 
of our DMA theory. 

To summarized, the property of excitable
AR networks has been discussed with the use of DMA.
The obtained results are summarized as follows.
(1) Depending on model parameters of $a$, $w$, $D$ and $N$,
AR networks display three types of dynamics (Fig. 1):
S, P and R states are characterized by the quantities
of $\zeta$, $\delta \zeta$, $\nu$ and $\sigma$,
as summarized in Table 1. 
(2) The S-P transition is of the (continuous)
second-order one while
P-R and S-R transitions are of the 
(discontinuous) first-order one 
with no hysteresis.
(3) There are no enhancements in order-parameter
fluctuations of $\delta \zeta$
at the transitions.
(4) Fluctuations in local variables ($\gamma$) remain finite
even for $N =\infty$ whereas those ($\rho$) in global variables
varies as $\rho \propto 1/N$, which is consistent with the
central-limit theorem.

\section*{Acknowledgements}
%The author would like to express his sincere thanks to
%Professor Hideo Nitta for critical reading of the manuscript.
This work is partly supported by
a Grant-in-Aid for Scientific Research from the Japanese 
Ministry of Education, Culture, Sports, Science and Technology.  

\newpage
\appendix
\section{Derivation of Eqs. (11)-(13)}

When we adopt the Gaussian decoupling approximation
given by Eqs. (10),
Eq. (9) becomes
\begin{eqnarray}
\frac{d \mu}{dt}
&=& \frac{1}{N} \sum_i \sum_{n=0}^{\infty}
\frac{F^{(2n)}}{(2n)!} B_{2n} <\delta \phi_i^2>^n \nonumber \\
&&+ \frac{w}{N^2} \sum_i \sum_j \sum_{n=0}^{\infty}
\frac{G^{(2n)}}{(2n)!}B_{2n} <(\delta \phi_j-\delta \phi_i)^2>^n,
\end{eqnarray}
where $B_{2n}=(2n-1)(2n-3)....3\cdot1$.
In deriving Eq. (A1), we treat $(\delta \phi_j-\delta \phi_j)$
as a new variable with the Gaussian distribution.
Adopting the mean-field approximation given by
\begin{eqnarray}
<\delta \phi_i^2>^n &\simeq & \gamma^{n-1} <\delta \phi_i^2>,\\
<(\delta \phi_i-\delta \phi_j)^2>^{n} &\simeq&
2^{n-1}\;(\gamma-\rho)^{n-1}
\;(<\delta \phi_i^2>+<\delta \phi_j^2>
-2 <\delta \phi_i \; \delta \phi_j>).
\hspace{0.2cm} \mbox{$(i \neq j)$}
%<\delta \phi_i^2> &\simeq& \gamma, \\
%\sum_{j}<\delta \phi_i \delta \phi_j> 
%&\simeq& \rho,
\end{eqnarray}
we get
\begin{eqnarray}
\frac{d \mu}{dt}
&=& \sum_{n=0}^{\infty}
\frac{F^{(2n)}}{2^n\; n!} \gamma^n
+ w\;\sum_{n=0}^{\infty}
\frac{G^{(2n)}}{n!} (\gamma-\rho)^n,
\end{eqnarray}
which yields Eq. (11).

From Eqs. (8) and (9), we get
\begin{eqnarray}
\frac{d \delta \phi_i}{dt}
&=& \frac{d \phi_i}{dt}-\frac{d \mu}{dt}, \\
&=& \sum_{n=0}^{\infty} F^{(2n+1)}
\frac{(\delta \phi_i)^{2n+1}}{(2n+1)!} 
+ \sum_{n=0}^{\infty}  F^{(2n)} 
[\frac{\delta \phi^{2n}}{(2n)!} - \frac{\gamma^n}{2^n\;n!}] \nonumber \\
&&+ \frac{w}{N}\sum_j
\sum_{n=0}^{\infty} G^{(2n+1)}
\frac{(\delta \phi_j-\delta \phi_i)^{2n+1}}{(2n+1)!} \nonumber\\
&&+ \frac{w}{N}\sum_j
\sum_{n=0}^{\infty}  G^{(2n)} 
[\frac{(\delta \phi_j-\delta\phi_i)^{2n}}{(2n)!} 
- \frac{(\gamma-\rho)^n}{n!}] + \xi_i(t).
\end{eqnarray}
With the use of Eq. (A6),
the calculation of $d \gamma/dt$ 
is performed as follows.
\begin{eqnarray}
\frac{d \gamma}{dt}&=& \frac{2}{N} \sum_i 
<\delta \phi_i \frac{d \delta \phi_i}{dt}>,  \nonumber\\
&=& \frac{2}{N} \sum_i \sum_{n=0}^{\infty}
\frac{F^{(2n+1)}}{(2n+1)!} <\delta\phi_i^{2n+2}> 
+ \frac{2}{N} \sum_i <\delta\phi_i\:\xi_i>  \nonumber\\
&& - \frac{2w}{N^2} \sum_i \sum_k 
\sum_{n=0}^{\infty} \frac{G^{(2n+1)}}{(2n+1)!}
<\delta \phi_i(\delta\phi_i-\delta\phi_k)^{2n+1}>, \\
&=& \frac{2}{N} \sum_i \sum_{n=0}^{\infty}
\frac{F^{(2n+1)}}{(2n+1)!} B_{2n+2}<\delta\phi_i^{2}>^{n+1}+2D  \nonumber\\
&& - \frac{2w}{N^2} \sum_i \sum_k 
\sum_{n=0}^{\infty} \frac{G^{(2n+1)}}{(2n+1)!}
B_{2n+2}<\delta \phi_i(\delta\phi_i-\delta\phi_k)>^{n+1}.
\end{eqnarray}
By using the mean-field approximation given by Eqs. (A2) and (A3) and
\begin{equation}
<\delta \phi_i(\delta\phi_i-\delta \phi_j)>^{n+1}
\simeq (\gamma-\rho)^n \;
(<\delta \phi_i^2>-<\delta \phi_j>),
\hspace{2cm} \mbox{$(i \neq j)$}
\end{equation}
we get
\begin{eqnarray}
\frac{d \gamma}{dt}&=& 2 \sum_{n=0}^{\infty} \frac{F^{(2n+1)}}{2^n\;n!}
\gamma^{n+1}
- 2 \sum_{n=0}^{\infty} 
\frac{G^{(2n+1)}}{n!} (\gamma-\rho)^{n+1} +2D,
\end{eqnarray}
which leads to Eq. (12).

The calculation of $d \rho/\d t$ is similarly performed by
\begin{equation}
\frac{d \rho}{dt}= \frac{1}{N^2} \sum_i \sum_j 
<\delta \phi_i \frac{d \delta \phi_j}{dt}
+\frac{d \delta \phi_i}{dt} \delta \phi_j>, 
\end{equation}
which yields Eq. (13).

For the AR model given by Eq. (14)
with $F(x)=1-a \:sin(x)$ and $G(x)=sin(x)$, we get 
\begin{eqnarray}
F^{(\ell)}(\mu)&=&c - a\;sin(\mu), \hspace{3cm}\mbox{($\ell=0$)}  \nonumber\\
&=& (-1)^{n+1} a\;sin(\mu), \hspace{2cm}\mbox{($\ell=2n > 0$)}  \nonumber\\
&=& (-1)^{n+1} a\;cos(\mu), \hspace{2cm}\mbox{($\ell=2n+1$)} \\
G^{(\ell)}(0)&=& 0, \hspace{5cm}\mbox{($\ell=2n$)}  \nonumber\\
&=& (-1)^{n}, \hspace{4cm}\mbox{($\ell=2n+1$)} 
\end{eqnarray}
which yield Eqs. (15)-(17).

%\newpage
\section{Derivation of Eqs. (54) and (55)}

The moment method takes into account means, variances and covariances
defined by
\begin{eqnarray}
m_i&=&<\phi_i>, \\
C_{ij}&=& <\Delta \phi_i \;\Delta \phi_j>,
\hspace{3cm} \mbox{($i,j=1$ to $N$)}
\end{eqnarray}
where $\Delta \phi_i=\phi_i - m_i$ and $C_{ii}$ denotes variances.
By adopting the Gaussian decoupling approximation
given by Eq. (10), we get
\begin{eqnarray}
\frac{d m_i}{d t}&=&\sum_{n=0}^{\infty} 
\frac{F_i^{(2n)}}{(2n)!}\;B_{2n} <\Delta \phi_i^2>
+ \frac{w}{N} \sum_k \sum_{n=0}^{\infty} 
\frac{G^{(2n)}}{(2n)!}\; B_{2n}
[<(\Delta \phi_k -\Delta \phi_i)^2>]^n, \\
\frac{d C_{ij}}{dt}&=& \sum_{n=0}^{\infty}
\frac{F_i^{(2n+1)}}{(2n+1)!}\; B_{2n+2}
(<\Delta\phi_j^2>^n+<\Delta \phi_i^2>^n)
\;<\Delta \phi_i \Delta \phi_j> \nonumber \\ 
&&+\frac{w}{N} \sum_k \sum_{n=0}^{\infty}
\frac{G^{(2n+1)}}{(2n+1)!} B_{2n+2}
[<(\Delta\phi_k-\Delta \phi_j)^2>^n\; 
<\Delta\phi_i(\Delta\phi_k-\Delta\phi_j)> \nonumber \\
&&+<(\Delta\phi_k-\Delta \phi_i)^2>^n\;
<\Delta\phi_j(\Delta\phi_k-\Delta\phi_i)>]
+ 2 D\; \delta_{ij},
\end{eqnarray}
where $F_i^{(\ell)}=F^{(\ell)}(m_i)$ and 
$G_i^{(\ell)}=G^{(\ell)}(0)$.
By a proper re-arrangement, Eqs. (B3) and (B4)
reduce to Eqs. (54)-(55).

%\newpage
\begin{center}
\begin{tabular}{|c||c|c|c|c|}
%\begin{tabular}{|c||*{4}{c|}}
\hline
type of states   &$\;\;\; \zeta \;\;\;$ & $\;\;\;\delta \zeta \;\;\;$
& $\;\;\; \nu \;\;\;$  &  $\;\;\; \sigma \;\;\;$
\\  \hline\hline
S     &    F    &     0   &   0   &    F   \\  \hline
R     &    0    &     0   &   F   &    0   \\  \hline
P     &    F    &     F   &   F   &    F   \\
\hline
\end{tabular}
\end{center}

\vspace{0.5cm}

{\bf Table 1} Quantities in the S, R and P states
of coupled AR networks: F and 0
denote finite and vanishing values, respectively.

%\begin{references}

%\end{references}

\begin{figure}
\caption{
%Fig A. 
The phase diagram of coupled AR networks,
showing the stationary (S) state, the time-periodic
(P) state and the random, diordered (R) state, calculated by DMA
with $N=10$ (thin solid curves), $N=100$ (solid curves)
and $N=\infty$ (dashed curves).
The dotted curve denotes the boundary obtained by SK (Ref.[7]).
Calculations by changing a parameter
of $D$ ($a$) along the horizontal (vertical) chain curve,
are presented in Fig. 2 (Fig. 3). 
}
\label{fig1}
\end{figure}

\begin{figure}
\caption{
%Fig. H. 
The $D$ dependence of (a) $\zeta$ and $\delta \zeta$, and
(b) $\nu$ and $\sigma$, for $a=1.05$, $w=1.0$ and $N=100$:
solid curves denote results calculated with the use of DMA: 
circles ($\zeta$), diamonds ($\delta \zeta \times 10$), 
squares ($\nu \times 10$) and triangles ($\sigma$)
express results obtained by simulations, dashed curves being drawn 
only for a guide of the eye.
}
\label{fig2}
\end{figure}

\begin{figure}
\caption{
%Fig. H. 
The $a$ dependence of (a) $\zeta$ and $\delta \zeta$, and
(b) $\nu$ and $\sigma$, for $D=0.1$, $w=1.0$ and $N=100$:
solid curves denote results calculated with the use of DMA: 
circles ($\zeta$), diamonds ($\delta \zeta \times 10$ ), 
squares ($\nu \times 10$) and triangles ($\sigma$)
express results obtained by simulations, dashed curves being drawn 
only for a guide of the eye.
}
\label{fig3}
\end{figure}

\begin{figure}
\caption{
%Fig. C. 
The distribution of local [$P(\phi(t))$] and
global variables [$P(\Phi(t))$] 
for $D=0.05$ [(a) and (b)], 
$D=0.10$ [(c) and (d)] and $D=0.30$ [(e) and (f)]
(in arbitrary units).
Dashed curves express simulation results.
Dotted curves in (a), (c) and (e) denote the
results of FPE for $N=\infty$.
}
\label{fig4}
\end{figure}

\begin{figure}
\caption{
%Fig. G.
(a) The time evolution of $P(\phi(t))$ calculated by DMA
for $a=1.05$, $w=1.0$, $D=0.10$ and $N=100$, and
(b) the time evolution of $n(\phi,t)$ calculated by FPE
for $a=1.05$, $w=1.0$, $D=0.10$ and $N=\infty$
(in arbitrary units).
}
\label{fig5}
\end{figure}

\begin{figure}
\caption{
%Fig. K.
The time evolution of (a) $\mid z(t) \mid$ and (b) $s(t)$
for $a=1.05$, $w=1.0$, $D=0.10$ and $N=100$:
solid and dashed curve denotes the results of DMA
and simulations, respectively.
}
\label{fig6}
\end{figure}

\begin{figure}
\caption{
%Fig. J. 
The $D$ dependence of (a) $\zeta$ and $\delta \zeta$, and
(b) $\nu$ and $\sigma$, for $a=1.05$, $w=1.0$ and $N=10000$
calculated with the use of DMA.
}
\label{fig7}
\end{figure}

\begin{figure}
\caption{
%Fig. F
The log-log plot of the $N$ dependence of
$\gamma$, $\rho$ and $\sigma$ for 
(a) $a=1.05$ and $D=0.05$ and (b) $a=1.20$ and $D=0.10$,
with $w=1.0$ and $N=100$. 
Solid curves denote results of DMA, and
Circles ($\gamma$), squares ($\rho$) and triangles ($\sigma$)
express those of simulations, dashed curves being only for
a guide of the eye. 
Right vertical scales are for $\sigma$ only.
}
\label{fig8}
\end{figure}

\begin{figure}
\caption{
%Fig. F
Distributions of output ISIs, $T_o$, 
as a function of $D$
for (a) $g=0$, (b) $g=0.1$ and $g=0.2$,
with $a=1.05$ and $N=100$.
Arrows in (a) denote
the S-P and P-R transition points.
}
\label{fig9}
\end{figure}

\end{document}